\documentclass[a4paper]{article}

\title{Integration of GRACE and PYTHIA
\footnote{Presented by S. Odaka at ACAT2000, Fermilab, Oct. 16 - 20, 2000.}}
\author{K. Sato,\(^{a} \) S. Tsuno,\(^{a} \) 
J. Fujimoto,\(^{b} \) T. Ishikawa,\(^{b} \) 
Y. Kurihara\(^{b} \) \\
and S. Odaka\(^{b} \)\footnote{E-mail: shigeru.odaka@kek.jp.}\\
\(^{a} \)Institute of Physics, University of Tsukuba, Tsukuba,
Ibaraki 305-8571, Japan \\
\(^{b} \)High Energy Accelerator Research Organization (KEK), \\
Tsukuba, Ibaraki 305-0801, Japan
}
\date{}

\begin{document}

\maketitle

\begin{abstract}
We have successfully developed a technique to integrate an automatic event-generator generation system GRACE and a general-purpose event generator framework PYTHIA. The codes generated by GRACE are embedded in PYTHIA in the created event generator program. The embedded codes give information on parton-level hard interactions directly to PYTHIA. The choice of PDF is controlled by the ordinary parameter setting in PYTHIA. This technique enables us to create easy-to-handle event generators for any processes in hadron collisions. Especially, in virtue of large capability of GRACE, we can easily deal with those processes containing many (four or more) partons in the final state, such as multiple heavy particle productions. This project is being carried out as a collaboration between the Japanese Atlas group and the Minami-Tateya group, aiming at developing event generators for Tevatron and LHC experiments.\\
\end{abstract}

GRACE \cite{grace} is a software package for automatic calculation in high energy physics, 
developed by the Minami-Tateya group.
The core part of the package is a program to generate Feynman diagrams 
relevant to specified initial and final states.
It then generates FORTRAN codes for calculating the corresponding cross section 
on the basis of the amplitude of each diagram.
The GRACE package also includes an integration and event-generation program 
called BASES/SPRING \cite{bases-spring}.
Hence, it provides us with a very powerful environment for developing 
event generators for studies of high energy physics.

Since the basic building block is the amplitude, 
GRACE has an advantage in those processes composed of many coherent diagrams.
(The number of elements increases quadratically if the calculation is based on 
the matrix element.)
This turns out to be an advantage in multi-body production processes, 
since in general the number of contributing diagrams increases as the number of final-state particles increases.
The program {\tt grc4f}\cite{grc4f} is a good example.
This is known to be the most reliable event generator at the tree level for four-fermion 
productions in electron-positron collisions at LEP2 energies.

In future hadron-collision experiments such as LHC,
interesting processes (Higgs-boson productions, SUSY-particle productions etc.)
in many cases result in multiple production of heavy particles ($Z/W$ and/or top quark) 
or cascade decays.
They produce many particles in the final state, 
and are composed of many coherent diagrams.
For example, we have to take into account 144 diagrams in total to evaluate the process 
$pp \rightarrow H^{0}b\bar{b} + X \rightarrow b\bar{b}b\bar{b} + X$.
Detailed studies of such processes require an exact (coherent) treatment of the whole 
production and decay reactions.
GRACE is expected to be a powerful tool for such studies.

However, GRACE can treat perturbative hard-scattering reactions only.
We need to add non-perturbative partonic structures of hadrons (PDF) 
and QCD evolutions in the initial and final states, 
in order to construct realistic event generators.
The most straightforward way to accomplish it is to connect GRACE to 
a general purpose event generator, such as PYTHIA \cite{pythia}, 
ISAJET \cite{isajet} and HERWIG \cite{herwig}.
We chose PYTHIA in this work.

There would be two ways for the connection.
One way is to interface them using an external data file, in which hard-scattering 
event data provided by GRACE are described.
Such a system can be flexible and portable.
The coding may be easier since only the I/O routines are necessary to be coded.
However, since the generation procedure is separated to two steps,
a special care or an appropriate software assistance is necessary to keep the consistency 
in the parameters.
This method is used in GRAPE \cite{grape}, a previous work of the Minami-Tateya 
group for electron-proton collisions.
A similar method is also adopted in the CompHEP for hadron collisions 
\cite{comphep-pythia}.

The second way is to embed the GRACE output codes in PYTHIA.
Although the coding may be more complicated, 
the event generation can be a single-step job in this method.
Since necessary parameters can be set through the parameter setting in PYTHIA, 
any inconsistency in the parameter choice can be automatically avoided.

We chose the latter method in the present work.
All the program codes describing the hard scattering are embedded in the subroutine 
{\tt PYUPEV}, prepared by PYTHIA to install user-defined processes.
This subroutine is called during the event generation 
if it is requested using {\tt PYUPIN} in the initialization stage.

For the event generation, it is important to give an appropriate definition of "kinematics", 
the mapping of kinematical variables to a set of uniform random numbers.
We have tried two methods for it.
In the first method, the mapping is fully defined by user-defined analytic functions.
The CPU time may be saved if the functions are appropriately defined.
The integration/event generation package BASES/SPRING is fully utilized 
in the second method, where a variable-grid mapping is implemented.
Since the grids are optimized by BASES, users need not to care about very details of 
the mapping.
Of course, in both methods, users need to choose an appropriate set of variables 
to avoid non-diagonal singularities.

The subroutine {\tt PYUPEV} that we have coded can be separated to two parts, 
an initialization stage and an event generation stage.
The initialization stage is called by {\tt PYUPIN} once in each job. 
The maximum of {\tt SIGEV} (see below) is searched to set 
the return argument {\tt SIGMAX} in the functional-mapping method.
In the grid-mapping method, BASES is called to estimate the total cross section
and {\tt SIGMAX} is set to be equal to it.
The grids are optimized here. 

PYTHIA calls the second stage in the event-generation loop with a frequency 
corresponding to {\tt SIGMAX}.
In this stage, first of all, one of the "sub-processes" is chosen.
In most cases in hadron collisions, every "process" of interest is composed of several 
incoherent sub-processes, 
since hadrons are composite and parton species in the final state are hard to identify.

The following procedure is different for two different methods of "kinematics".
In the functional-mapping method, a set of kinematical variables defining an event is 
determined from a set of uniform random numbers according to the defined "kinematics".
The differential cross section of this event is calculated using the GRACE output codes.
The return argument {\tt SIGEV} is calculated from the differential cross section, PDF 
and the Jacobian of the "kinematics".

Instead, the event generator SPRING is called in the grid-mapping method.
SPRING generates an event using the GRACE output codes, PDF and the grid information 
optimized by BASES.
{\tt SIGEV} is always equal to {\tt SIGMAX}.

In both methods, the above procedure is followed by  a Lorentz boost to the laboratory frame 
and a determination of the color flow.
It should be noted that the color flow can be determined automatically using the information 
from BASES.

After returning to PYTHIA, an event sampling is done according to the weight 
{\tt SIGEV/SIGMAX}.
The sampling is dummy (i.e., all events are accepted) in the grid-mapping method.
After that, the initial- and final-state parton radiations are added by PYTHIA to simulate 
the effect of QCD evolution, resulting in an underlying hadronic activity and a finite transverse momentum of the hard-scattering system.

The coding has been refined by developing actual event generators.
So far, we have developed generators for the following processes:
\begin{equation}
  pp({\rm or} p\bar{p}) \rightarrow q \gamma + X,
  \label{process1}
\end{equation}
\begin{equation}
  pp({\rm or} p\bar{p}) \rightarrow W g + X \rightarrow \mu \nu g +X,
  \label{process2}
\end{equation}
\begin{equation}
  pp({\rm or} p\bar{p}) \rightarrow H^{0}W + X \rightarrow b \bar{b} \mu \nu +X,
  \label{process3}
\end{equation}
\begin{equation}
  pp({\rm or} p\bar{p}) \rightarrow b \bar{b}{\rm (QCD)}W + X 
  \rightarrow b \bar{b} \mu \nu +X,
  \label{process4}
\end{equation}
\begin{equation}
  pp({\rm or} p\bar{p}) \rightarrow H^{0} b \bar{b} + X 
  \rightarrow b \bar{b} b \bar{b} +X,
  \label{process5}
\end{equation}
\begin{equation}
  pp({\rm or} p\bar{p}) \rightarrow b \bar{b} b \bar{b}{\rm (QCD)} + X. 
  \label{process6}
\end{equation}
The results were compared with those from other existing generators, 
PYTHIA built-in generators and CompHEP, to examine the coding.
We found reasonable agreements in all cases.

As an example, the performance of a developed event generator is compared with 
that of a PYTHIA built-in generator in Table~\ref{table1} for one of the simplest cases, 
Process~(\ref{process2}).
In this study, the weak boson was not made to decay and the simulations of both initial- 
and final-state parton radiations were turned off.
The two "kinematics" methods were tried.

\begin{table}
  \begin{tabular}{lccc}
    \hline
    &Functional mapping &Grid mapping &PYTHIA({\tt ISUB=16}) \\
    \hline
    Total cross section (nb)&63.36$\pm$ 0.20 &63.43$\pm$ 0.13 &63.17$\pm$ 0.20 \\
    Generation efficiency (\%) &19 &35 &19  \\
    CPU time for 100k events (min) &12.5 &20.3 &4.6  \\
    \hline
  \end{tabular}
  \caption{Performance of a developed event generator for the process 
  $pp \rightarrow W g + X$ ($\sqrt{s} = 14~{\rm TeV}, p_{T}(g) \geq 5~{\rm GeV}/c$). 
  The results are compared with those of a PYTHIA built-in generator. 
  A Linux-PC with 300 MHz-Pentium II is used. 
  The weak boson is not made to decay. The initial- and final-state parton radiations are 
  not simulated. The full simulation consumes another 45 minutes for 100k events.}
  \label{table1}
\end{table}

The cross section is in very good agreement, showing that the integration of 
GRACE and PYTHIA is done successfully.
The functional-mapping method gives a better performance in CPU time 
than the grid-mapping method, as expected.
Although both GRACE + PYTHIA generators consume appreciably longer CPU time 
than the PYTHIA built-in generator, 
the difference is not very serious since the simulation of parton radiations and hadronization, 
which are not implemented in this study and common to all three generators, 
takes much longer time.

The most noticeable advantage of the GRACE + PYTHIA system is in the fact
that using this technique we can easily develop event generators for those processes 
which are not and/or hard to be implemented in PYTHIA.
Process~(\ref{process5}) is one of such processes.
The total cross section estimated by using the developed generator is shown 
in Table~\ref{table2} for three assumed Higgs-boson masses, 
and compared with the result of CompHEP. 
We have applied the grid-mapping method only in this case.
The agreement with CompHEP is quite good.

\begin{table}
  \begin{tabular}{ccc}
    \hline
    $M(H^{0})$ (GeV) &GRACE (fb) &CompHEP (fb) \\
    \hline
    80&6.006 &6.083 \\
    120 &0.989 &1.002 \\
    160 &0.357 &0.356  \\
    \hline
  \end{tabular}
  \caption{Comparison between GRACE and CompHEP for the process 
  $p\bar{p} \rightarrow H^{0}b\bar{b} + X \rightarrow b\bar{b}b\bar{b} + X$ 
  at $\sqrt{s} = 2~{\rm TeV}$.
  The total cross section is evaluated for three cases of the Higgs-boson mass.}
  \label{table2}
\end{table}

We plan to make some improvements in order to make the development easier.
So far, routines interfacing PYTHIA and GRACE are written by hand.
Since the functionality of these routines are almost common to all generators, 
we will be able to make them automatically generated by changing the libraries referred 
by GRACE.

Processes in hadron collisions are in most cases composed of 
several incoherent sub-processes.
The present GRACE system cannot handle such processes automatically.
Some modifications and additions by hand are necessary now. 
We would like to automate these tasks.

The difference between the sub-processes is, in most cases, the difference 
in quark species in the initial and/or final states.
If we can treat quark masses and their couplings as variables, 
these sub-processes can share an identical code for the calculation.
The Cabbibo-Kobayashi-Masukawa matrix can be implemented automatically 
if such a treatment is realized.

In summary, 
we have established a technique to embed GRACE output codes in PYTHIA.
This technique allows us to develop new hadron-collision event generators easily.
We have applied the technique to some two-, three- and four-body production processes.
Obtained results are in good agreement with the results from existing generators.
The generator for four b-quark productions, Processes (\ref{process5}) and 
(\ref{process6}), will be released in a few months.
Since GRACE is expected to be advantageous in multi-body production processes,
we would like to go to five- or six-body production processes as the next step.
We also have an automatic next-to-leading order (NLO) calculations 
for hadron-collision processes in our view.

This project is being carried out as a collaboration between the ATLAS-Japan 
group composed of Japanese members of the ATLAS collaboration at LHC, 
and the Minami-Tateya group,
aiming at developing event generators for Tevatron and LHC.
K.S., S.T. and S.O. are from the ATLAS-Japan group, 
and J.F., T.I. and Y.K. from the Minami-Tateya group. 
There are many other people from these two groups who have contributed 
to this work.
Among them, the authors wish to thank here Y. Takaiwa, S. Kawabata and K. Kato 
for their educational contributions, and T. Abe for discussions.


\begin{thebibliography}{1}

\bibitem{grace}
T. Ishikawa et al., GRACE manual, KEK Report 92-19 (1993);
F. Yuasa et al., Prog. Theor. Phys. Suppl. 138, 18 (2000); hep-ph/0007053. 

\bibitem{bases-spring}
S. Kawabata, Comput. Phys. Commun. 41, 127 (1986);
Comput. Phys. Commun. 88, 309 (1995).

\bibitem{grc4f}
J. Fujimoto et al., Comput. Phys. Commun. 100, 128 (1997).

\bibitem{pythia}
T. Sjostrand, Comput. Phys. Commun. 82, 74 (1994).
We used an updated version (6.138) for the present work.

\bibitem{isajet}
H. Baer, F.E. Paige, S.D. Protopopescu and X. Tata, hep-ph/0001086.

\bibitem{herwig}
G. Marchesini et al., Comput. Phys. Commun. 67, 465 (1992);
G. Corcella et al., hep-ph/9912396.

\bibitem{grape}
T. Abe, hep-ph/0012029, to appear in Comput. Phys. Commun.;
T. Abe et al., Proc. Workshop on Monte Carlo Generators for HERA Physics 
(Plenary Starting Meeting), DESY-PROC-1999-02 (1999) p. 566.

\bibitem{comphep-pythia}
V.A. Ilyin, A.E. Pukhov and A.N. Skachkova, talk in this workshop;
E.E. Boos et al., Proc. the Second Int. Workshop on Software Engineering, 
ed. D. Perret-Gallix (World Scientific, Singapore, 1992) p. 665.

\end{thebibliography}
\end{document}